\documentclass[aps,prl,twocolumn,preprintnumbers,showpacs,nofootinbib]{revtex4}

\usepackage{graphicx}
\usepackage{dcolumn}
\usepackage{bm}
\begin{document}
\newcommand{\Od}{{\cal O}}
\newcommand{\lsim}   {\mathrel{\mathop{\kern 0pt \rlap
  {\raise.2ex\hbox{$<$}}}
  \lower.9ex\hbox{\kern-.190em $\sim$}}}
\newcommand{\gsim}   {\mathrel{\mathop{\kern 0pt \rlap
  {\raise.2ex\hbox{$>$}}}
  \lower.9ex\hbox{\kern-.190em $\sim$}}}


\title{The nature of branon dark matter}

\author{A.L. Maroto}
\affiliation{Departamento de  F\'{\i}sica Te\'orica,
 Universidad Complutense de
  Madrid, 28040 Madrid, Spain}%

\date{\today}

\begin{abstract}
We explore the different possibilities
for branons as dark matter candidates. We consider 
a general brane-world model, parametrized by the number of
extra dimensions $N$, the fundamental scale of gravity $M_D$, the brane
tension scale $f$ and the branon mass $M$. We analyse the parameter region in which
branons behave as collisionless thermal relics (WIMPs), either cold or hot (warm) 
together with less standard scenarios in which they are strongly self interacting or 
produced non-thermally.
 \\
\end{abstract}

\maketitle

\section{Introduction}
Cold dark matter (CDM) is a fundamental ingredient of the current cosmological standard  
model.  
An enormous variety of observations at very large scales ($\gsim 1$ Mpc), from  cosmic
microwave background anisotropies, galaxy surveys, cluster abundances or Ly-$\alpha$
forest  are successfully explained within this framework. However, to achieve these
goals, the dark matter 
component is required to behave as a collisionless non-relativistic particle fluid. 
This poses an important  problem from the particle physics point of view. 
Indeed, no candidate with the assumed properties exists 
within the known particles, and therefore, physics beyond the 
Standard Model appears as the only feasible solution. 
Among the proposed candidates, we find, on one hand,  the axion which is the Goldstone
boson associated to the spontaneous breaking of the Peccei-Quinn symmetry postulated
to solve the strong CP problem of QCD. The production  of axions in the early
universe  mainly takes place through the so called misalignment mechanism in which
the $\Theta$ angle is initially displaced from its
equilibrium value  $\Theta=0$ and oscillates coherently. Such oscillations can be
intrepreted as a zero-momentum Bose-Einstein condensate which essentially behaves
as a non-relativistic matter fluid. Despite the fact that axions are light
particles, this non-thermal 
mechanism produces  cosmologically important energy densities. 
 On the other hand we have the thermal relics, produced by the
well-known freeze-out mechanism in an expanding universe. They are typically 
weakly interacting massive
particles (WIMPs) such as the neutralino in supersymmetric theories \cite{kam}
(for a 
recent review see also \cite{carlos}).
In addition to their weak interactions with SM particles, these candidates
usually also have very weak self-interactions (collisionless).

Despite the success of CDM at large scales, the model exhibits certain 
difficulties at sub-galactic scales. In particular, high resolution N-body simulations
of dark halos show cuspy density profiles \cite{NFW,Moore} which contradict 
observations from low
surface brightness galaxies \cite{deBlok} and dwarfs \cite{Moorea,Flores} 
that suggest flat density profiles. In addition,
CDM also predicts too many small subhalos within  simulated larger systems 
\cite{Klypin, Moorec}, in 
contradiction with observations of the number of satellite galaxies in the Local
Group. Solutions which aim to reduce the power 
at small scales, 
but keeping the good properties of CDM at large scales, have 
been proposed at both the astrophysical
 and a more fundamental level. They include modifications
of the primordial power spectrum at small scales \cite{liddle}, or of the collapse histories of
CDM and baryons \cite{bullock}, but also  modifications in the nature of the dark matter. 
There are two main proposals along the latter lines, namely,
warm dark matter (WDM) \cite{dolgov} and self-interacting dark matter (SDM)
\cite{spergel}. In the WDM scenario
the particle dark matter is still collisionless, but much lighter 
with $M\lsim 1$ keV. Free-streaming exponentially suppresses  the power spectrum
below the scale 
$\lambda_{FS}\simeq 0.2\, (\Omega_{Br}h^2)^{1/3}
(\mbox{keV}/M)^{4/3}\,\mbox{Mpc}$. This suppression solves the sub-halo problem 
provided $M\simeq 1$ keV \cite{hogan}, however the central cusp problem requires a much lighter
particle $M\lsim 300$ eV \cite{hogan}. In addition, the Ly-$\alpha$ forest in quasar
spectra observations puts a severe
lower bound on the mass of the WDM candidate, namely, $M\gsim 750$ eV, which 
makes the simplest models unfeasible \cite{narayanan}. A second possibility is to take into
account dark matter
self-interactions. Thus, if the mean free path of dark matter particles
is of the order of $\sim 1$ kpc, but still keeping small annihilation cross sections, then
heat can be transported out of the center of the halo smoothing out the central
cusp, which now would become a more spherical core. 
In order for this mechanism to work, the total elastic dark matter cross section
should be in the range \cite{dave}:
\begin{eqnarray}
\frac{\sigma}{M}\sim \,0.5 - 5 \,\,\mbox{cm$^2$ g$^{-1}$}
\end{eqnarray}  
which is comparable to a nucleon-nucleon interaction.

However, the SDM model is also very strongly constrained by different observations. Thus,
in order to avoid evaporation of halos on a Hubble timescale, the cross
section should be sufficiently small $\sigma/M < 0.3 - 1$ cm$^2$ g$^{-1}$ \cite{gnedin}. 
In addition Chandra observations of the mass density profile of a certain cluster imposes 
a very strong limit $\sigma/M < 0.1$ cm$^2$ g$^{-1}$ \cite{yoshida}. 
Finally, the elliptical shape 
of a particular cluster determined by lensing data leads to 
a even stronger limit $\sigma/M < 0.02$ cm$^2$ g$^{-1}$ \cite{miralda}, otherwise
self-interactions would erase ellipticity. The interpretation of these results
together with other direct constraints can be found in \cite{markevitch}. 
It is also interesting to point out that the combination of both scenarios, i.e. 
self-interacting warm dark matter, could alleviate their individual problems 
\cite{Hannestad}. Apart from SDM and WDM,  an interesting proposal in
\cite{Lin} shows that non-thermal WIMP production could also help resolving the
mentioned discrepancies.

The new requirements imposed by these models on the dark matter particles makes 
it even more difficult to
find appropriate candidates in the mentioned SM extensions. For that reason it 
is worth analysing
the potential of branon dark matter \cite{CDM} as an alternative to the more common neutralino
or axion cases. In this paper we study the different possibilities in terms of the
brane-world parameters. We will consider  the case in which branons
and SM particles are the only relevant degrees of freedom at low
energies. This happens whenever the brane tension scale $f$ is much smaller
than the fundamental scale of gravity in $D=4+N$ dimensions $M_D$
\cite{bando}. 

The paper
is organized as follows: after introducing branon dark matter and its interactions, we
study its collisional (collisionless) nature essentially in terms of $M_D$ and $f$. 
Then we analyse the possibility of producing
branons non-thermally, in a similar way to the misalignment mechanism, where now
the brane coherently oscillates along the extra dimensions. We end
with some conclusions.

\section{Branon interactions}
In the case in which traslational invariance is an exact symmetry in the
extra dimensions, the presence of the brane can break it spontaneously, and 
branons can be identified with the corresponding Goldstone bosons
\cite{sundrum,doma}. 
For the sake of definiteness, we will consider a
particular model with three compact extra dimensions, although we expect that
the results will depend only on the different energy scales
involved in the problem and not on the details of the model.
We will choose an  extra space $B=S^3$, so that its isometry group 
$G(B)=SU(2)\times SU(2)$ is spontaneously broken down to $SU(2)$. Therefore, the
coset space is $K=SU(2)\times SU(2)/SU(2)=SU(2)\sim S^3=B$. We thus have three branons, 
and 
the coset space metric is nothing but the 3-sphere metric: 
$h_{\alpha\beta}(\pi)=\delta_{\alpha\beta}+\pi^\alpha\pi^\beta/(v^2-\pi^2)$, 
where $v=f^2R_B$ with $R_B=M_D^{-1}(M_P/M_D)^{2/N}$ the size of the extra space $B$
and $M_P$  the usual Planck mass. Therefore, 
$v$ is the spontaneous symmetry
breaking scale which sets the size of the coset space $K$. 

As shown in \cite{doma},
the low-energy branon dynamics is given, to lowest order in
derivatives, by the non-linear sigma model corresponding 
to that
symmetry breaking pattern, i.e.:
\begin{eqnarray}
{\cal L}_{Br}&=&
\frac{1}{2}g^{\mu\nu}h_{\alpha\beta}(\pi)\partial_{\mu}\pi^\alpha\partial_\nu\pi^\beta
+\Od(p^4)\nonumber
\\
&\simeq& \frac{1}{2}g^{\mu\nu}\partial_{\mu}\pi^\alpha
\partial_{\nu}\pi^\alpha+\frac{1}{2v^2}g^{\mu\nu}
\pi^\alpha\pi^\beta\partial_{\mu}\pi^\alpha
\partial_{\nu}\pi^\beta\nonumber \\
&+&\Od(\pi^6)+\Od(p^4)
\label{lag}
\end{eqnarray}
We see that $v$ sets the strength of branon self-interactions to lowest order
in derivatives. Notice however that the $\Od(p^4)$ terms are not suppressed by the $v$
scale, but by the $f$ scale \cite{doma}.

So far we have only considered massless branons, however, traslational 
invariance in the extra dimensions can be explicitly broken, and
branons can acquire mass.  In such a case, we should include in the Lagrangian
the corresponding breaking terms. As shown in \cite{sky,collider}, such terms will include
an arbitrary number of even powers of branon fields, i.e.:
\begin{eqnarray}
{\cal L}_{break}&=&-\frac{M^2}{2}\pi^2+\frac{\lambda}{4!}\frac{M^2}{v^2}\pi^4+\Od(\pi^6) 
\label{break}
\end{eqnarray}  
Since branons are pseudoscalar particles, parity on the brane forbids terms with an odd 
number of fields \cite{cosmo}.
Notice that here we are assuming for simplicity that the branons are degenerate, with a 
mass $M$ which
is small compared to $v$. Therefore the explicit breaking terms can be also organized in
powers of the branon mass, which is considered to be of the same order
as a derivative in the power counting. Accordingly, we expect the 
parameter $\lambda$, which depends on the specific breaking mechanism,  
to be of order one. This is completely analogous to the pion dynamics in Chiral 
Perturbation Theory \cite{libro}, in which $v$ would play the role of the
pion decay constant and the expression in (\ref{break}) with $\lambda=1$ 
corresponds to the expansion
in $\pi$ fields of the lowest order mass term. 

Branons also interact with the SM particles through their energy momentum tensor.
Again the lowest order term in derivatives was obtained in \cite{doma,collider}: 
\begin{eqnarray}
{\cal L}_{Br-SM}&=& \frac{1}{8f^4}(4\partial_{\mu}\pi^\alpha
\partial_{\nu}\pi^\alpha-M^2\pi^\alpha\pi^\alpha g_{\mu\nu}+\dots)
T^{\mu\nu}_{SM}\label{lag}\nonumber \\
\end{eqnarray}
where the dots stand for higher order terms  in $\pi$ fields.  We see that 
the branon-SM interactions are controlled by the brane tension scale $f$. 
What is important from the viewpoint of the dark matter problem is that branon 
self interactions are controlled by a completely different energy scale from
branon-SM interactions. This means that it is possible to have strong
self-interactions and simultaneously small annihilation cross sections into SM particles, 
so that the branon relic abundance can be important. In addition, since branons always interact 
by pairs this implies that they are stable particles and therefore they fulfill all the
requirements of a dark matter candidate.

\section{Collisional vs. collisionless branons}
Branons were in thermal equilibrium with the rest
of SM particles when the temperature of the universe was above 
their decoupling temperature, which
depends on $f$ and $M$ through the corresponding annihilation cross section. 
Assuming that the evolution of the universe was standard up to a temperature
of the order of the brane tension scale, and that decoupling occured
below such temperature, which is indeed the case (see \cite{CDM}), then the above 
effective Lagrangians 
can be used to compute the branon thermal relic abundances generated by 
the standard freeze-out mechanism. 
The calculation of the abundances \cite{CDM,cosmo} show that there are two regimes 
in which the corresponding energy density could be
compatible with the measurements  of the cosmological dark matter, i.e. 
$\Omega_{Br}h^2\lsim 0.129$ and also with LEP-II bounds \cite{collider}. On one hand we have a
region in which $M\gsim 100$ GeV and $f\lsim M$, which corresponds to cold relics, 
and on the other, a
region with $M\lsim 180$ eV and $f\gsim$ 200 GeV in which branons behave as 
hot or warm dark matter.

In this section we explore these two regimes and calculate the branon self-interactions
in terms of the number of branon fields $N$ and the fundamental scale $M_D$. With
that purpose we will obtain that branon-branon elastic cross section. 

Since particles in the halo collide with very low relative velocities 
$v_{rel}\sim 10^{-3}$, 
the scattering amplitudes can be obtained from the above effective Lagrangian 
and 
they are given, for $\lambda=1$,  
by the well-known Weinberg's low-energy theorem \cite{libro}:
\begin{eqnarray}
A(s,t,u)=\frac{s-M^2}{v^2}
\end{eqnarray}
Assuming that the explicit breaking terms
respect the remaining $SU(2)$ symmetry, branons transform as an $SU(2)$ triplet and
 it is possible to construct isospin $I=0,1,2$ amplitudes as:
\begin{eqnarray}
T_0(s,t,u)&=&3A(s,t,u)+A(t,s,u)+A(u,t,s)\nonumber \\
T_1(s,t,u)&=&A(t,s,u)-A(u,t,s)\nonumber \\
T_2(s,t,u)&=&A(t,s,u)+A(u,t,s)
\end{eqnarray}
and also define the corresponding partial wave amplitudes with definite
angular momentum $J$:
\begin{eqnarray}
t_{IJ}=\frac{1}{32\pi}\int_{-1}^1 d(\cos \theta) P_J(\cos \theta) T_I(s,t,u)
\end{eqnarray}
From the amplitude, the corresponding total elastic scattering cross section
summed over final isospin states and averaged over initial ones
in the non-relativistic limit reads:
\begin{eqnarray}
\sigma_{tot}= \frac{23M^2}{384\pi v^4}+\Od(v_{rel}^2)
\label{sigma}
\end{eqnarray}
Apparently this cross section grows unboundedly with $M$, this is due to the
fact that we are working with an effective theory, and tree-level
unitarity violations are expected for sufficiently high values of 
$s\simeq 4M^2$. In order to 
estimate the range of validity of the above result, 
we will use the well-known
unitarity limit:
\begin{eqnarray}
\mbox{Re}\; t_{IJ}\leq \frac{1}{2\rho} 
\end{eqnarray} 
where $\rho=(1-4M^2/s)^{1/2}$. From the dominant $t_{00}$ wave, we get
in the mentioned limit:
\begin{eqnarray}
\frac{M^2}{v^2}\leq \frac{16\,\pi}{7\;v_{rel}}
\end{eqnarray}
which is compatible with the general result derived in \cite{Hui}. 
For  typical values of $v_{rel}$, we find $M/v\lsim \Od(10^2)$.
Substituting back in (\ref{sigma}), we find a limit for the cross section:
\begin{eqnarray}
\frac{\sigma_{tot}}{M}\lsim \frac {4 \cdot 10^6}{M^3}
\end{eqnarray}

In order for this kind of branons to be strongly self-interacting they should satisfy 
$\sigma_{tot}/M \gsim 0.5$ cm$^2$ g$^{-1}$=
$2.3 \cdot 10^{3}$ GeV$^{-3}$, which implies $M\lsim 12$ GeV, in agreement
with the general bound \cite{Hui}. This inmediately
excludes cold branons as possible SDM candidates, according to the LEP-II limits
mentioned before. 
Notice that, in any case, this is only a unitarity limit on the model, for
larger masses, branons interactions could be sufficiently strong, but they will
not be described by our effective theory.

\begin{figure}[h]
\resizebox{8.5cm}{!}{\includegraphics{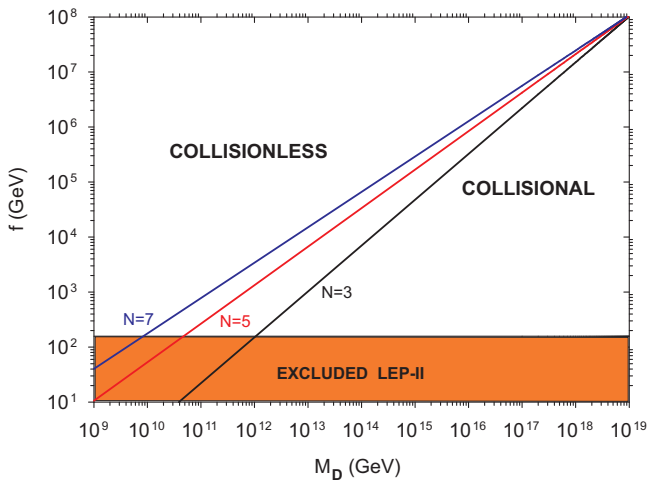}}
\caption{Collisional vs. collisionless regions in $f-M_D$ space for 
a model with $M=100$ eV and N=3,5,7. The dividing lines correspond
to $\sigma_{tot}/M\simeq 0.5$ cm$^2$ g$^{-1}$. The lower striped area is excluded by
LEP-II single photon analysis.}
\end{figure}

Nevertheless, self-interacting warm branons with masses in the region $M\lsim 180$ eV 
are not a priori excluded 
by the previous bound. In principle, such a light particle would conflict with 
Ly-$\alpha$ forest observations, as commented before. However, such limits were derived
assuming collisionless WDM. In the collisional case  free-streaming is suppressed and the relevant
cut-off scale for structure formation is the standard Jeans length $\lambda_J$, which can be 
substantially smaller than $\lambda_{FS}$ \cite{hogan}. In any case a detailed calculation
of the matter power spectrum would be needed for each particular model \cite{Hannestad}. 

In Fig. 1  we have plotted the different regions in the $f-M_D$ plane for a branon mass
 $M\sim 100$ eV. The curves are not very sensitive to the branon mass, since
they scale as $M^{1/8}$. We see that for a given $f$, models become self-interacting at
sufficiently high $M_D$. Notice that the plotted regions are consistent with the
assumption of low-tension brane $f\ll M_D$.

\section{Thermal vs. non-thermal branons}
So far we have considered branons as WIMPs which were produced thermally
in the early universe. However, if the evolution of the universe was not standard
before branon decoupling or the reheat temperature $T_{RH}$ was
sufficiently low, branons were never in
thermal equilibrium with radiation.
However, still they could be produced non-thermally very much in the
same way as axions \cite{axions}. Thus, if the maximum temperature reached in the 
universe was smaller than the freeze-out temperature $T_{RH}\ll T_f$, but
larger compared to the explicit symmetry breaking scale $T_{RH}\gg \Lambda$
with $\Lambda=(Mv)^{1/2}$, then brane fluctuations were initially
essentially massless and decoupled from the rest of matter fields.
In this case, there is no reason to expect that the position in the
extra dimension $Y_0$ at which the brane is created should coincide with
the minimum of the branon potential ($Y=0$). In general we expect 
$Y_0\sim \Od(R_B)$, i.e. $\pi_0\sim v$ within a region of size $H^{-1}$
\cite{axions}. The evolution of the branon
field is then simply that of a scalar field in an expanding universe. Thus, 
while $H\gg M$, the field remains frozen in its initial position
$\pi=\pi_0$. Below the temperature $T_i$ for which $3H(T_i)\simeq M$,
the branon field oscillates around the minimum.
These oscillations correspond to a zero-momentum
branon condensate, its energy density behaving like non-relativistic matter.

For the sake of simplicity we will consider the case of branons which are
lighter than all the SM particles masses, i.e. we will assume that $M\ll 1$ MeV and that
neutrinos are massless. In such a case the branon condensate cannot decay by
particle production into massive SM particles, but still branons could annihilate
through $2\rightarrow 2$ processes into massless gauge bosons or neutrinos. 
Since the expectation value of $\pi$ varies on horizon scales, the 
typical momentum of branons can be estimated as $\vert \vec p\vert\sim H$, i.e.
$\vert \vec p\vert\sim M$ at the beginning of oscillations (if there was a previous period
of inflation then the typical momentum would be much smaller). It is 
possible to estimate the annihilation rate from the results in \cite{cosmo}
as $\Gamma_{2\rightarrow 2}\sim M^7v^2/f^8$ at the first stages
of oscillations. In the regions in which we will
be interested, we find $\Gamma_{2\rightarrow 2}\ll M$.   
In addition,  those processes in which
four zero-momentum branons are transformed into two branons with energies
$E=2M$ were already considered in the first reference in \cite{axions}.
In this case Bose enhancement effects are important.
The results, obtained for axions, can be inmediately translated 
into the branons case, simply
identifying $v$ with the Peccei-Quinn scale. It was
shown that in the initial stages of oscillations, which corresponds to $\pi\sim v$,
the decay rate was $\Gamma_{4\pi\rightarrow 2\pi}\sim M$. Again, this
rate would be smaller if there was a period of inflation. Therefore
the necessary condition to avoid the energy depletion of  the branon condensate 
is $H(T_{RH})\gsim M$. Making use of the Friedmann equation in a radiation dominated
universe, this implies $T_{RH}\gsim (MM_P)^{1/2}$, which automatically ensures 
$T_{RH}\gg \Lambda$.
 Since $\Gamma$
decreases very fast due to the universe expansion, then it will never become comparable
to $H$. 
Accordingly, the branon condensate energy
density essentially is not reduced by particle production, 
but only diluted by the Hubble 
expansion. We see that the reheating temperature
should satisfy the condition $T_i\simeq (M M_P)^{1/2}< T_{RH}< T_f$ in order for the
condensate to form and survive until present. Therefore if $T_i> T_f$
the above interval disappears and only 
thermal relics can be present. In the opposite case, $T_i<T_f$, we can also
have non-thermal production. In the case of light branons, a good estimation
for the freeze-out temperature was obtained in \cite{CDM,cosmo}: 
$\log (T_f/\mbox{GeV})\simeq (8/7)\log (f/\mbox{GeV})-3.2$.
In Fig. 2 the $T_i=T_f$ curve which separates the thermal and non-thermal
regions is plotted as a dashed line. We see that for sufficiently light or weakly
coupled branons the non-thermal production is possible.

\begin{figure}[h]
\begin{center}
\resizebox{8.5cm}{!}{\includegraphics{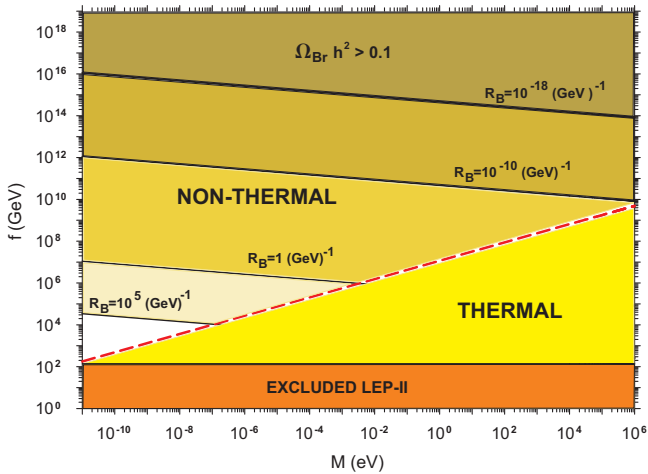}}
\caption{Thermal vs. non-thermal branons regions in the $f-M$ plane. 
The dashed (red) line separates the 
two regions and corresponds to $T_i\simeq T_f$. The continuous (black) lines correspond to
$\Omega_{Br}h^2\simeq 0.1$ for different values of the radius of the extra dimensions
$R_B=10^5,\, 1,\,10^{-10},\, 10^{-18}$ GeV$^{-1}$.}
\end{center}
\end{figure}

Following the same steps as in the axionic case, we can
 calculate the energy density which is stored today in the form of 
branon oscillations. Assuming that $M$ does not depend on the temperature, it
is given by \cite{axions}:
\begin{eqnarray}
\Omega_{Br}h^2\simeq \frac{2.5 N\,v^2\,M\, T_0^3}{M_P\,T_i\,\rho_0}
\simeq \frac{6.5\cdot 10^{-20}N}{\mbox{GeV}^{5/2}} f^4\,R_B^2\,M^{1/2}\nonumber \\
\end{eqnarray}
where $T_0$ and $\rho_0$ are the photon temperature and critical density today
respectively.

In particular it can be seen that within certain parameters regions, the 
above energy density can be cosmologically important. Thus 
in Fig. 2 we plot the $\Omega_{Br}h^2=0.1$ curves, corresponding to the 
current observational values,  for different values of $R_B$
in the $f-M$ plane. Thus the region above each curve would be excluded by non-thermal
branon overproduction. This is another manifestation of the cosmological
moduli problem. Notice that in the allowed non-thermal regions of the figure the 
condition $f\ll M_D$ is also satisfied. 

In the case in which the branon masses are larger than some SM particle mass, then
the corresponding annihilation channel would open up and a detailed analysis
of the  energy loss rate would be necessary. In any case,  the
possibility of having heavier non-thermal branons is not a priori excluded.

\section{Conclusions}
We have explored the different possibilities of branon dark matter in terms of the
brane-world parameters $(N,M_D,f,M)$. We have shown that
apart from the standard
scenario \cite{CDM} in which branons behave as collisionless thermal relics, the
large parameter space of these models allows for branons to behave also as collisional or 
non-thermal dark matter candidates. 
In the collisional case, this opens an interesting possibility since
the existing proposals for SDM candidates in the literature are very limited 
\cite{Qballs,Hannestad}. Nevertheless, the allowed parameters region 
reduces only to warm collisional branons  and therefore 
a detailed analysis of the actual effect on structure formation is needed.


 {\bf Acknowledgements:} I would like to thank J.A.R. Cembranos, A. Dobado and J.R. Pel\'aez
for useful comments and discussions, and also Alberto L. del Amo for additional motivation.
This work
 has been partially supported by the DGICYT (Spain) under the
 project numbers FPA 2000-0956 and BFM2002-01003.

\end{document}